\documentclass{aa}  

\usepackage{graphicx}
\usepackage{mathtools} 
\usepackage{relsize}
\usepackage{multirow}
\usepackage{multicol}
\usepackage{booktabs}
\usepackage{subfigure}

\usepackage{siunitx}
\usepackage{natbib}
\bibpunct{(}{)}{;}{a}{}{,} % to follow the A&A style
\usepackage{txfonts}
\usepackage[colorlinks,citecolor=blue]{hyperref}
\begin{document}

   \title{Radio multiwavelength analysis of the compact disk CX Tau: \\Presence of strong free-free variability or anomalous \\microwave emission}

   \author{P.~Curone
          \inst{1},
          L.~Testi
          \inst{2, 3},
          E.~Macías
          \inst{4},
          M.~Tazzari
          \inst{3},
          S.~Facchini
          \inst{1},
          J.~P.~Williams
          \inst{5},
          C.~J.~Clarke
          \inst{6},
          A.~Natta
          \inst{7},
          G.~Rosotti
          \inst{1, 8},
          C.~Toci
          \inst{4},
          G.~Lodato
          \inst{1}
          }

   \institute{Dipartimento di Fisica, Università degli Studi di Milano, via Celoria 16, 20133 Milano, Italy\\
              \email{pietro.curone@unimi.it}
        \and
             Alma Mater Studiorum Università di Bologna, Dipartimento di Fisica e Astronomia (DIFA), Via Gobetti 93/2, 40129 Bologna, Italy     
        \and
            INAF-Osservatorio Astrofisico di Arcetri, L.go E. Fermi 5, I-50125 Firenze, Italy
        \and
            European Southern Observatory, Karl-Schwarzschild-Str. 2, 85748 Garching bei München, Germany
        \and
            Institute for Astronomy, University of Hawaii, Honolulu, HI 96822, USA
        \and
            Institute of Astronomy, University of Cambridge, Madingley Road, CB3 0HA Cambridge, UK
        \and
            School of Cosmic Physics, Dublin Institute for Advanced Studies, 31 Fitzwilliams Place, Dublin 2, Ireland
        \and
            Leiden Observatory, Leiden University, P.O. Box 9513, 2300 RA Leiden, The Netherlands
         }

   \date{Received 29 May 2023 / Accepted 17 July 2023}
 
  \abstract
   {Protoplanetary disks emit radiation across a broad range of wavelengths, requiring a multiwavelength approach to fully understand their physical mechanisms and how they form planets. Observations at submillimeter to centimeter wavelengths can provide insights into the thermal emission from dust, free-free emission from ionized gas, and possible gyro-synchrotron emission from the stellar magnetosphere. This work is focused on CX Tau, a ${\sim}0.4\,M_\odot$ star with an extended gas emission and a compact and apparently structureless dust disk, with an average millimeter flux compared to Class II sources in Taurus. We present observations from the Karl G. Jansky Very Large Array (VLA)  across  four bands (between  \SI{9.0}{\mm} and  \SI{6.0}{\cm}) and combine them with archival data from the Atacama Large Millimeter/submillimeter Array (ALMA), the Submillimeter Array (SMA), and the Plateau de Bure Interferometer (PdBI). This multiwavelength approach allows us to separate the dust continuum from other emissions. After isolating the dust thermal emission, we derived an upper limit of the dust disk extent at \SI{1.3}{\cm,} which is consistent with theoretical predictions of a radial drift-dominated disk. The centimeter data show a peculiar behavior: deep observations at \SI{6.0}{\cm} did not detect the source, while at \SI{1.3}{\cm,} the flux density is anomalously higher than adjacent bands. Intraband spectral indices suggest a dominant contribution from free-free emission, whereas gyro-synchrotron emission is excluded. To explain these observations, we propose a strong variability among the free-free emission with timescales shorter than a month. Another possible interpretation is the presence of anomalous microwave emission from spinning dust~grains.}

   \titlerunning{Multiwavelength analysis of CX~Tau}
   \authorrunning{P. Curone et al.}
   \maketitle

%
%-------------------------------------------------------------------

\section{Introduction}
\label{sec:introduction}

Protoplanetary disks are complex systems consisting of gas, dust, and ionized particles surrounding young stars, and they emit radiation over a wide range of wavelengths from radio to X-rays. A multiwavelength approach therefore becomes crucial to understanding the physical mechanisms occurring in disks and explain how they evolve and possibly go on to form planets.  

\begin{figure*}[]
\centering
\includegraphics[width=\hsize]{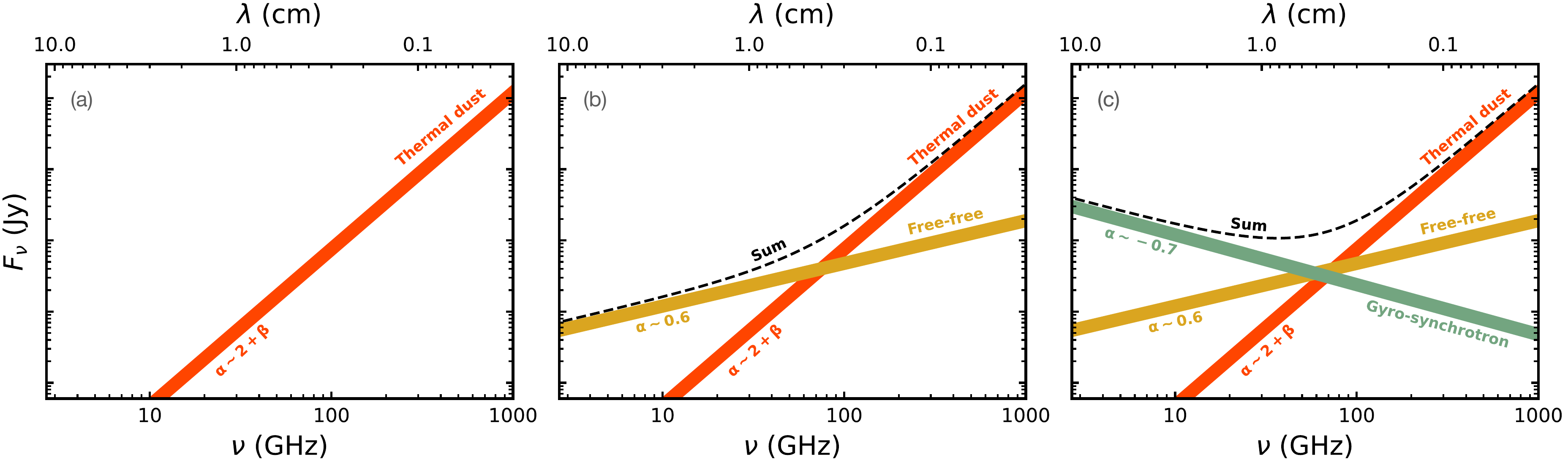}
\caption{Illustrative plot for the main components in the continuum emission of a typical protoplanetary disk in the wavelength range between the submillimeter to the centimeter. The three panels are described in Sect.~{\ref{sec:introduction}}.}
\label{fig:multiwav_ppdisks_sketch}
\end{figure*}

In the last few decades, observations in the submillimeter to centimeter wavelength range revealed different emission mechanisms by analyzing their spectral flux density distributions (e.g., \citealt{Rodmann2006, Ricci10, Pascucci2014,  Sheehan2016, Coutens2019}). In this context, a powerful diagnostic able to disentangle the various physical origins of the emission is the spectral index, $\alpha = \log_{10}\left[ F_{\nu_1} / F_{\nu_2}\right] \, / \, \log_{10}[\nu_1 / \nu_2]$, where $F_{\nu_1}$ and $F_{\nu_2}$ are the flux densities measured in the observing wavelengths, $\nu_1$ and $\nu_2$, respectively. We  summarize the most common cases in Fig.~\ref{fig:multiwav_ppdisks_sketch}, based on the dominant physical mechanism of emission. The first category (panel~a) consists of disks in which the thermal emission from dust dominates at all wavelengths. Assuming optically thin emission and Rayleigh-Jeans approximation, dust emission follows a spectral index of $\alpha{\sim}2+\beta$, where $0\lesssim \beta \lesssim2$ is the dust opacity spectral index and depends on the maximum grain size (see, e.g., \citealt{Draine2006, Ricci10, Testi2014, Tazzari21_multiwav}). The second category (panel~b) includes disks in which dust emission dominates down to a wavelength of a few millimeters, followed by a change in the spectral index due to the contribution of free-free emission at longer wavelengths. Free-free emission originates through the interaction of free electrons with ions in the ionized gas present in the innermost regions of disks or in jets and its spectral index can range from -0.1 to 2.0 in the case of a totally optically thin or optically thick emission, respectively  \citep{Ubach2017}. A typical value is $\alpha {\sim} 0.6,$ as predicted by the theoretical models of an expanding partially optically thick spherical wind \citep{PanagiaFelli1975_freefree} or a collimated conical jet \citep{Reynolds1986}. Photoevaporative winds can also produce optically thin free-free emission with a spectral index of ${\sim}\,{-}0.1$ \citep{Pascucci2012}, and both jet and wind could coexist (e.g., \citealt{Macias2016}). Finally, the third category (panel~c) includes disks in which a gyro-synchrotron outburst is occurring and dominates at centimeter wavelengths. This non-thermal emission arises from the interaction of electrons with the stellar magnetosphere. Its spectral index depends on the electron energy distribution and it is generally negative with a typical value of $\alpha{\sim}\,{-}0.7$ \citep{Essential_Radio_Astronomy}.

The advantages of detailed multiwavelength analysis to study single protoplanetary disks have been explored in a few recent papers (\citealt{CarrascoGonzalez2019} for HL~Tau, \citealt{Macias2021} for TW~Hya, and \citealt{Guidi2022} for HD~163296). In this work, we focus on CX~Tau. It is a M2.5 star \citep{Luhman2018_spectraltype} located at a distance from the Sun of $126.5\pm0.3 \,\mathrm{pc}$, as estimated by \citet{2021AJ....161..147B}, using Gaia EDR3 \citep{GaiaCollaboration2021_EDR3}. \cite{Simon2019_Mdyn} measured a dynamical mass of $0.38\pm0.02\,M_\odot$ and \cite{Andrews2013} obtained a stellar bolometric luminosity $L_\star = 0.38\pm0.05\,L_\odot$ and an effective temperature $T_\mathrm{eff}=3475\pm130\,\mathrm{K}$. The estimated accretion rate is $7.1\times 10^{-10}\,M_\odot \ \mathrm{yr}^{-1}$ \citep{Hartmann1998_accretion}.  From the infrared excess in the SED, \cite{Najita2007} classified the source as a transitional disk, but the presence of an inner cavity has not been detected with high angular resolution observations by \cite{Facchini_19}.  CX~Tau hosts a compact dust disk and an extended gas disk, with a ratio between the gas and dust radii of 5.4 \citep{Facchini_19}, indicating that the disk evolution is dominated by dust radial drift \citep{Weidenschilling1977}. This is in contrast to the majority of disks that show a ratio between the gas and dust radii of ${\sim}2.5$, interpreted by invoking planet formation that is halting dust radial drift (\citealt{Sanchis2021}, \citealt{Toci_RCO}).

CX~Tau is an excellent candidate for a thorough multiwavelength analysis for two main reasons. First, it represents a typical protoplanetary disk, due to its small size along with a millimeter flux at the 50th percentile of the flux distribution of Class II disks in Taurus. Second, CX~Tau has already been observed by the Atacama Large Millimeter/submillimeter Array (ALMA) in Band~6 (\SI{1.3}{\mm}) at a high resolution of 0.04'' (corresponding to ${\sim} \SI{5}{\mathrm{au}}$),  and the results are presented in  \cite{Facchini_19}. At this angular resolution, there is no sign of substructures and the data are consistent with a smooth disk model. The authors estimated a total flux density of 9.75 mJy and a radius enclosing 68\% of the continuum flux density ($R_{68\%}$) equal to \SI{14}{\mathrm{au}}. Moreover, they revealed that the brightness radial profile of CX~Tau is consistent with the central regions of the profiles of DSHARP disks that are not in a binary system and do not present an inner cavity. Therefore, CX~Tau is not a scaled-down version of more extended disks but differs from the latter only for the lack of rings at large radii.

In this work, we present observations from the Karl G. Jansky Very Large Array (VLA) of CX~Tau in Ka (\SI{9.0}{\mm}), K (\SI{1.3}{\cm}), Ku (\SI{2.0}{\cm}), and C (\SI{6.0}{\cm}) bands. Combining these observations with archival data from ALMA in Band~7 (\SI{0.9}{\mm}) and with values from the literature in the (sub)millimeter, we obtained a multiwavelength view of CX~Tau. This paper is organized as follows. Section~\ref{sec:obs_data_reduction} presents the observations and the procedures we used to reduce and calibrate them, Sect.~\ref{sec:analysis_and_results} describes the analysis and the obtained results, and Sect. \ref{sec:discussion_and_conclusions} discusses these findings in light of multiwavelength analyses in general.

\section{Observations and data reduction}
\label{sec:obs_data_reduction}

CX~Tau was observed with VLA in the Ka band in B configuration (maximum baseline of \SI{11.1}{\km}) on 12 October 2020 (VLA Project 20A-373, PI: M. Tazzari) with an on-source integration time of 1.6~hours. The spectral windows covered the range between 29 and \SI{37}{\GHz} (corresponding to wavelengths of $8.1-\SI{10.3}{\mm}$). During the observation, 3C147 was used as bandpass and absolute flux calibrator, J0403+2600 as the pointing calibrator, and J0438+3004 as complex gain (amplitude and phase) calibrator.

Under the same VLA project, CX~Tau was observed in the Ku band in C configuration (maximum baseline of \SI{3.4}{\km}) on 2 and 4 March 2020 for a total on-source integration time of 24~minutes. The bandwidth extended from 12 to \SI{18}{\GHz} (wavelengths between 1.7 and $\SI{2.5}{\cm}$). The flux and band calibrator was 3C147, the complex gain calibrator was J0431+2037, and both sources were used to calibrate the pointing.

For the C band, we have observations from two different VLA projects. Under project 20A-373, observations were performed in C configuration on 20 February 2020 with a total time on-source of 15~minutes. The wavelength coverage was from 4 to \SI{8}{\GHz} ($3.7 - \SI{7.5}{\cm}$). For the calibration, the same sources as for the Ku band data were used. Then, CX~Tau was observed again in the C band but with an A configuration (maximum baseline of \SI{36.4}{\km}) between 9 and 11 January 2021, for a total integration time of 3.8 hours (VLA Project 20B-299, PI: M.~Tazzari). Here, 3C147 was used as bandpass and flux calibrator and J0403+2600 as complex gain calibrator.

Observations in the K band were performed more recently, on 9 April 2022 (VLA Project 22A-401, PI: M.~Tazzari). The integration time was 15~minutes and the spectral windows ranged from 18.4 to \SI{26.2}{\GHz} (wavelengths from 1.1 to \SI{1.6}{\cm}). 3C147 was employed for the calibration of pointing, bandpass, and flux, while J0403+2600 was the complex gain calibrator.

First, all these observations were calibrated by the VLA pipeline. Then, we performed the self-calibration using the software \texttt{CASA}, version 6.2 \citep{CASA2022}. We executed a spectral average on each dataset taking into account the requirements to avoid bandwidth smearing\footnote{\url{https://casadocs.readthedocs.io/en/v6.3.0/notebooks/synthesis_imaging.html}}, but did not employ time averaging during the self-calibration. For the imaging, we employed Briggs weighting with robust 1 as the best compromise between signal-to-noise ratio (S/N) and side lobes effects, and the  \texttt{mtmfs}  deconvolver to properly take into account the relevant bandwidth over observing frequency ratio in VLA data.  We performed one round of phase-only self-calibration for the observations in the Ka and Ku bands obtaining  improvements in the peak S/N by a factor of 2.3 and 2.9, respectively. For the C-band observation, we kept the observations from the two different configurations separate. In both cases,   we executed one round of phase self-calibration always taking as a reference the flux density from the other sources in the field of view, given the lack of a detection for CX~Tau emission at this wavelength. We obtained a modest  ${\sim}10\%$ increase in the S/N. We placed a particular emphasis on the treatment of the K-band data, carefully checking the calibration executed with the VLA pipeline and performing four rounds of phase-only self-calibration which highly improved the peak S/N by a factor of 7.8 (see Appendix~\ref{Kband_checks} for details).

In addition to these VLA observations, we also reduced and calibrated the ALMA archival data in Band~7 (\SI{0.9}{\mm}) from the project 2013.1.00426.S (PI: Yann Boehler) presented in \cite{Simon17}. We performed four rounds of phase-only self-calibration, for an improvement in the peak S/N of 20\%.

\section{Analysis and results}
\label{sec:analysis_and_results}

\subsection{Spectral flux density distribution}
\label{sed:SED}

\begin{figure}[]
\centering
\includegraphics[width=\hsize]{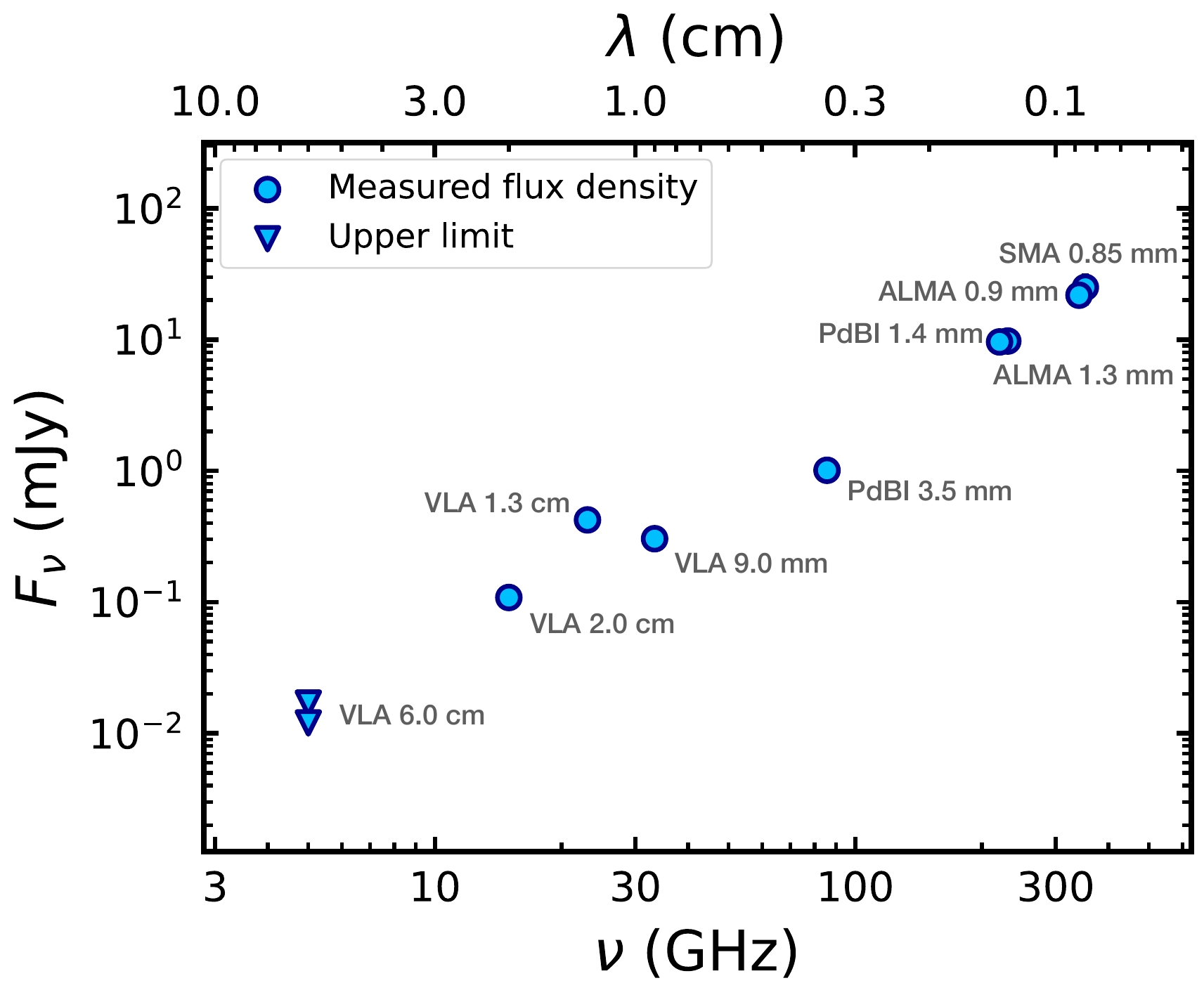}
\caption{Spectral flux density distribution of CX~Tau including data in the submillimeter to centimeter range. For each value of the flux density we report the total uncertainty including the statistical and systematic flux calibration errors. In most cases, the error bars fall within the plotted points. Non-detections are indicated by the upper limits, obtained as the flux from a visibility fit of a central point source plus $2\times\mathrm{RMS}$.}
\label{fig:CXTau_SED}
\end{figure}

Figure~\ref{fig:CXTau_SED} presents the spectral flux density distribution with all the observations in the submillimeter to centimeter range, including data from VLA, ALMA, the Submillimeter Array (SMA), and the Plateau de Bure Interferometer (PdBI) (\citealt{Andrews_Williams_2005}, \citealt{Ricci10}, \citealt{Pietu2014}, \citealt{Simon17}, \citealp{Facchini_19}). The ALMA and VLA images, along with a table containing properties from all the observations, are given in Appendix~\ref{app:images_table}.

To estimate the integrated flux density and, for the wavelengths where the source is resolved, the radial extent of the dust emission, we performed a characterization in the $uv$-plane using the code \texttt{galario} \citep{Tazzari2018_galario}. We computed the best-fit model using a Markov chain Monte Carlo (MCMC) approach, assuming that the emission is axisymmetric (see Appendix~\ref{app:galario_resuts}). The observations with ALMA at \SI{0.9}{\mm}  are spatially resolved, so we employed a Gaussian profile to model the intensity profile: $I(R) = f_0 \,\exp{(-R/2\sigma^2)}$, where $f_0$ is a normalization term, $\sigma$ characterizes the Gaussian centered in the disk center, and $R$ is the radial coordinate. The continuum emission in the VLA observations at \SI{9.0}{\mm} and \SI{2.0}{\cm} are unresolved, so we modeled the disk as a central point source  $I(R) = f_0 \,\delta (R)$, where $\delta(R)$ is the Dirac delta function. To perform this fit for the \SI{2.0}{\cm} data, we subtracted the CLEAN model visibilities of an external source in the field of view  at an angular distance of ${\sim} \SI{100}{\arcsec}$ from CX~Tau using the CASA task \texttt{uvsub}.  In every run, we also fit for the right ascension and declination offsets ($\Delta \mathrm{RA}$, $\Delta \mathrm{Dec}$), but we kept fixed the disk inclination and position angle from the estimates of \cite{Facchini_19} ($\mathrm{inc}=55.1^\circ$, $\mathrm{PA}=66.2^\circ$), as they were obtained with data that had a higher resolution and sensitivity than all  the other observations. In the two 6.0~cm observations, a clear central source was not detected. To estimate a flux density upper limit, we adopted the same procedure as \cite{Barenfeld2016}. We employed the \texttt{CASA} task \texttt{uvmodelfit} and fit a central point source to the visibilities of each dataset, after subtracting the other sources in the field of view with \texttt{uvsub}. To this value, we added $2\times\mathrm{RMS}$, so that the upper limit contains the real source flux with a confidence level of 95\%. At \SI{1.3}{\cm}, despite the highest angular resolution available with the VLA at this wavelength (${\sim} \SI{0.1}{\arcsec}$ corresponding to ${\sim}\SI{13}{\mathrm{au}}$), the source is dominated by an unresolved component but shows hints of a  marginally resolved source. To extract the flux density, we fit these data with a point source, but an attempt to retrieve the radial extent is described in Sect.~\ref{sec:R_vs_nu}. We present all the results of these \texttt{galario} fits in Appendix \ref{app:galario_resuts}.

\subsection{Evaluating time variability at centimeter wavelengths}
\label{sec:time_var}

\begin{figure*}[]
\centering
\includegraphics[width=\hsize]{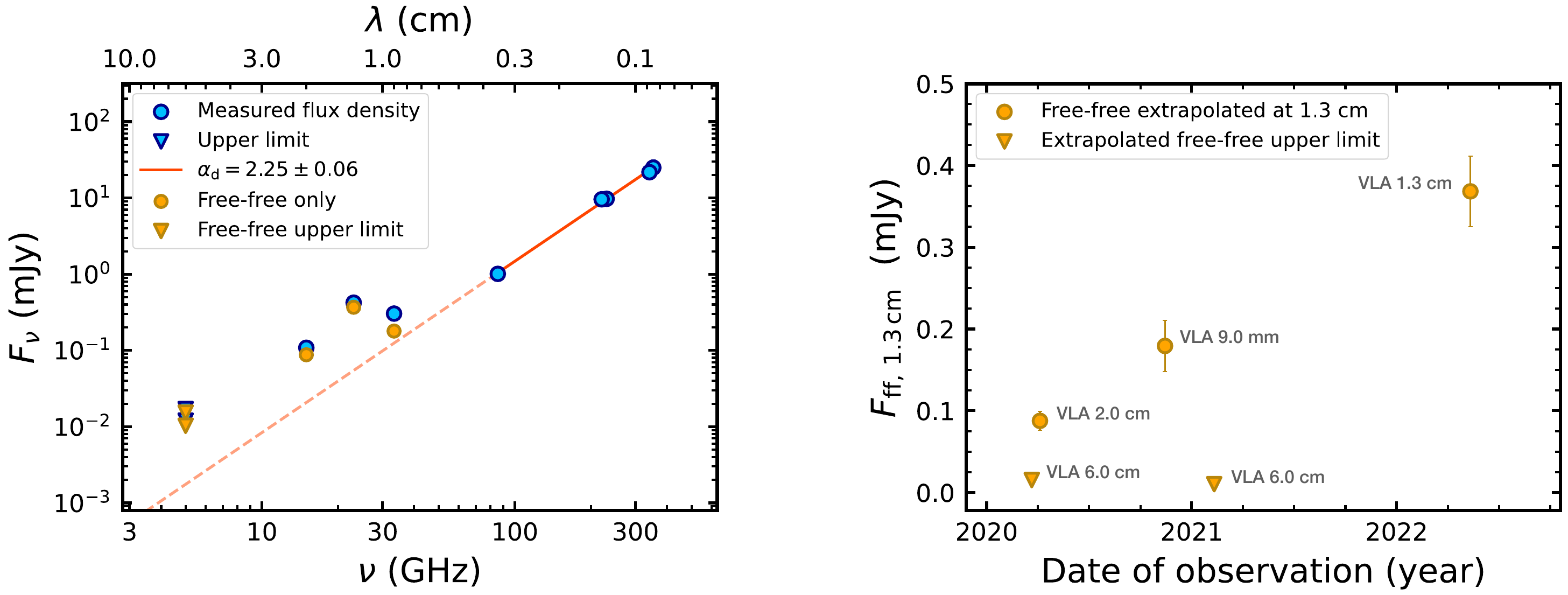}
\caption{Extraction and time variability of the free-free component. \textit{Left panel}: Flux density distribution of CX~Tau, showing the dust emission fit (orange line) used to extract the free-free only estimates (yellow circles and arrows) from the measured flux densities and upper limits. \textit{Right panel}: Time variability of extrapolated free-free estimates at 1.3~cm. The extrapolation was performed assuming a free-free spectral index of $0.7\pm0.1$. The values in the plot are labeled with the observation they  came from originally.}
\label{fig:SED_time_variability}
\end{figure*}

Looking at the spectral density distribution in Fig~\ref{fig:CXTau_SED}, two characteristics appear particularly surprising. First, the two observations at \SI{6.0}{\cm} did not detect any emissions, despite the stringent upper limit of \SI{12.0}{\mathrm{\mu Jy}} provided by the data taken in A~configuration. Second, the emission at \SI{1.3}{\cm} is significantly higher  than  the emissions in the adjacent bands at \SI{9.0}{\mm}  and \SI{2.0}{\cm}. In particular, the \SI{1.3}{\cm} observation has a flux density higher by a factor of ${\sim} 2$ than a power-law fit between the emission at \SI{9.0}{\mm} and \SI{2.0}{\cm} would suggest. Checking the intraband spectral indices of these three VLA detections resulted for each observation in a value between ${\sim} 0.5$ and 1.1 (see Appendix~\ref{app:intraband_spectral_indices}). This is compatible with the typical value of the free-free spectral index $\alpha{\sim}0.6$ \citep{PanagiaFelli1975_freefree, Reynolds1986}, while excluding contributions from gyro-synchrotron emission that are typically characterized by a negative spectral index. Therefore, we interpret the VLA observations as dominated by free-free emission and explain the non-detections at \SI{6.0}{\cm} and the anomalously high emission at \SI{1.3}{\cm} as due to the time variability of free-free.

The procedure we applied to evaluate the time variability is summarized in the left panel of Fig.~\ref{fig:SED_time_variability}. We first assumed that the flux density detected in the observations between \SI{0.9}{\mm} and \SI{3.5}{\mm} is dominated by dust emission only. Fitting a power-law to these data, we obtained a dust spectral index $\alpha_\mathrm{d}=2.25\pm0.06$ (orange solid line), consistent with  partially optically thick dust emission and/or grain growth in the assumption of optically thin emission and Rayleigh-Jeans approximation \citep{Draine2006}. Extrapolating the fit to centimeter wavelengths (orange dashed line) makes it apparent that the flux density from VLA detections diverges from millimeter data, in a way that resembles the behavior in panel b of Fig~\ref{fig:multiwav_ppdisks_sketch}. Then, assuming that VLA detections only have contributions from free-free and dust emission, we obtained the estimates of free-free alone (yellow circles) by subtracting  the extrapolation of the dust emission fit from the measured flux densities  for the respective observing wavelength. With the intent of comparing the free-free emissions at the same wavelength, we chose \SI{1.3}{\cm} as the reference and extrapolated to this wavelength the free-free only estimates at the other VLA bands, assuming a spectral index $\alpha_\mathrm{ff}=0.7\pm0.1$ consistent with the intraband spectral indices of the VLA detections. For the free-free extraction and extrapolation, the two upper limits at \SI{6.0}{\cm} are treated in the same way as the detections at \SI{9.0}{\mm}, \SI{1.3}{\cm}, and \SI{2.0}{\cm}. We report in the right panel of  Fig.~\ref{fig:SED_time_variability} the extrapolated values of the free-free only emission at \SI{1.3}{\cm} as a function of the time of their respective observations. There is significant variability both in amplitude and also in time, particularly noticeable by the fact that the first non-detection at \SI{6.0}{\cm} and the \SI{2.0}{\cm} measurement are separated by ${\sim}2$~weeks. It should be noted that we used a constant spectral index for dust emission at all wavelengths to be conservative. Typically, the dust spectral index steepens at centimeter wavelengths due to grain properties and size distribution \citep{Wilner2005}. In spite of this, a steeper spectral index would not affect the evident variability in the extrapolated free-free emission.

\subsection{Relation between dust radius and observing wavelength}
\label{sec:R_vs_nu}

Studying the dust radius dependence on observing wavelength provides a useful metric to assess disk evolution. Since different wavelengths are more sensitive to the emission of different dust grain sizes, we expect that the observed disk radius should change with the observing frequency. Specifically, the theoretical prediction by \cite{Rosotti19} (using dust evolution models that account for grain growth and drift by \citealt{Birnstiel12}) suggests that  at centimeter wavelengths, probing bigger grains subject to stronger radial drift \citep{Weidenschilling1977}, the  extent of a radial-drift-dominated disk should be smaller  compared to the radii measured at shorter wavelengths. 

\begin{figure}[]
\centering
\includegraphics[width=\hsize]{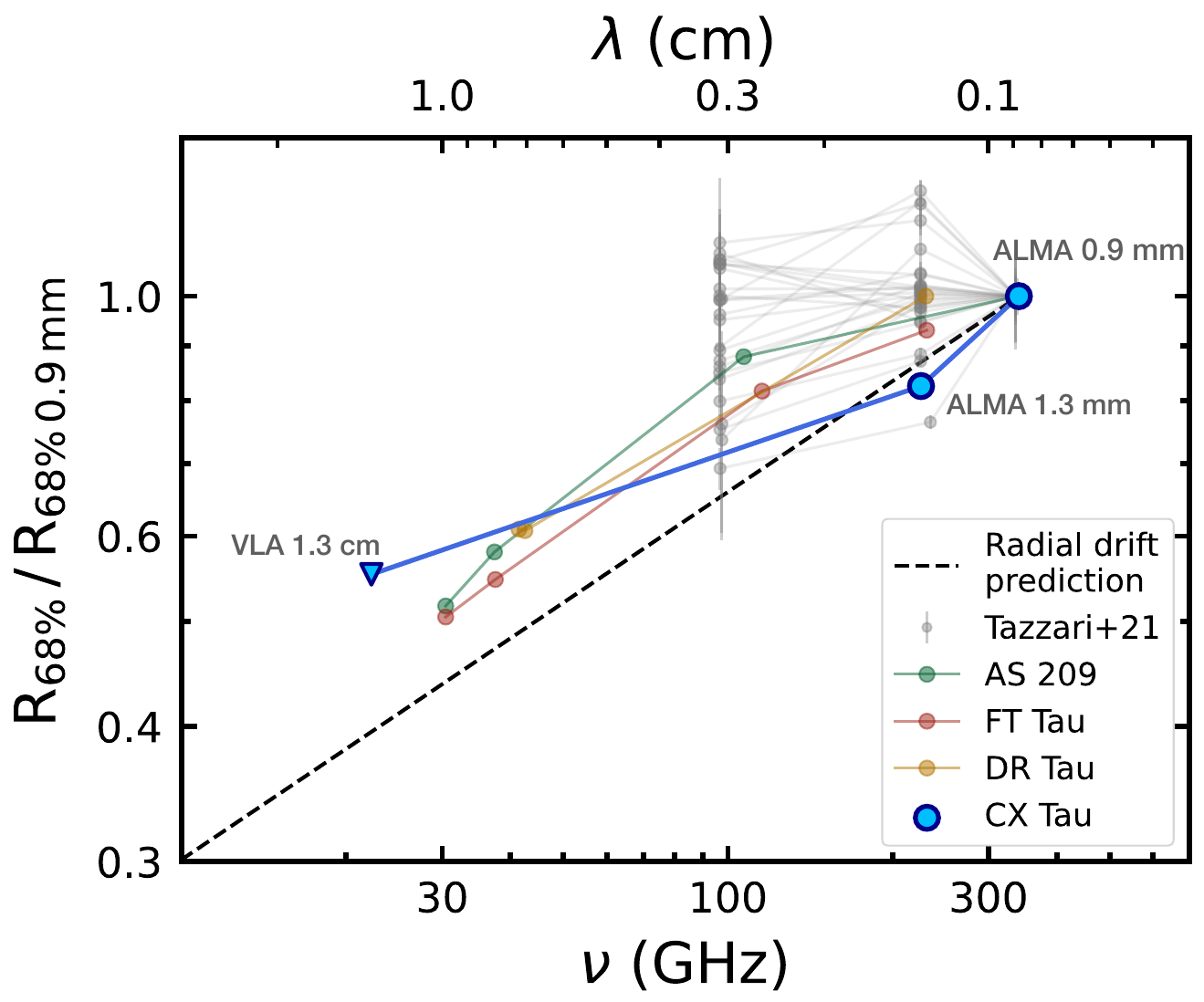}
\caption{Relation between the measured $R_{68\%}$ (normalized to the value at 0.9~mm) and the observing wavelength for CX~Tau, compared with the radial drift prediction by \cite{Rosotti19} and with the same relation for AS~209, FT~Tau, DR~Tau \citep{Perez2012_AS209, Tazzari2016}, and the disks of the Lupus survey by \cite{Tazzari21_multiwav}.}
\label{fig:R68_vs_freq}
\end{figure}

For CX~Tau, \cite{Facchini_19} at \SI{1.3}{mm} measured a radius enclosing 68\% of the flux density ($R_{68\%}$) of $14.0\pm0.3\,\mathrm{au}$. From the \texttt{galario} fit with a Gaussian profile of ALMA \SI{0.9}{\mm} data, we obtained $R_{68\%} = 16.9\pm0.1\,\mathrm{au}$. The VLA observation with the highest angular resolution is the one in the K band at \SI{1.3}{\cm} with a beam of ${\sim}\SI{0.1}{\arcsec}$. The source is mostly unresolved, but we retrieved an upper limit for the dust size by excluding the noisy visibilities at longer baselines and subtracting the estimate of the unresolved free-free only emission to the real part of all remaining visibilities (see details in Appendix~\ref{app:upper_lim_Kband}). Fitting the residual visibilities with \texttt{galario} using a Gaussian profile, we acquired $R_{68\%}=9.2\pm1.5\,\mathrm{au}$. We interpret this value as an upper limit because the subtraction of a constant value to the real part of all visibilities implies that  some unresolved dust emission has possibly been excluded, thereby skewing the inferred extent to a larger size. Figure~\ref{fig:R68_vs_freq} presents the relation between  the measured $R_{68\%}$ (normalized to the value at 0.9~mm) and the observing wavelength for CX~Tau compared to the radial drift prediction by \cite{Rosotti19}. The same relation is also shown for AS~209, FT~Tau, DR~Tau (data from \citealt{Perez2012_AS209, Tazzari2016}), and the disks within the survey in Lupus of \cite{Tazzari21_multiwav}. Unlike the other sources,  CX~Tau appears consistent with the theoretical prediction for  a disk dominated by radial drift. This is in line with the fact that CX~Tau is a rare example of a disk whose dust emission is smooth even with high-resolution ALMA observations (another example is PDS~66, \citealt{Ribas2023}). Moreover, the ratio between the gas and dust radii of 5.4 measured by \cite{Facchini_19} at \SI{1.3}{\mm} matches the results from the population synthesis study of \cite{Toci_RCO} for a \cite{Shakura_Sunyaev} disk viscosity value of ${\sim}10^{-3}-10^{-4}$.

\section{Discussion and conclusions}
\label{sec:discussion_and_conclusions}

In this work, we present a multiwavelength view of the compact protoplanetary disk CX~Tau in the submillimeter to centimeter range. After a careful subtraction of the contaminant emission, we detected dust thermal continuum at \SI{1.3}{\cm}, indicating the presence of large grains and the dominance of dust radial drift in the evolution of this disk. This is in line with the observed high ratio between the gas and dust radii consistent with radial drift (\citealt{Facchini_19}, \citealt{Sanchis2021}, \citealt{Toci_RCO}). Such a scenario implies that dust traps may not be forming within these disks -- or, alternatively, the planet formation process is restricted to the inner regions. In either case, CX Tau presents an intriguing example suggesting the need for future observations with higher resolution.

We explain the peculiar scatter of the VLA data in the flux density distribution as the result of a strong time variability of unresolved free-free emission. Free-free emission is known to rarely vary in intensity by more than ${\sim}30\%$ over a timescale of some weeks to a few months (e.g., \citealt{Sheehan2016}, \citealt{Coutens2019}). CX~Tau, instead, shows a rapid variability of about two weeks associated with a change in intensity by a factor of ${\sim}20$ (see the difference in the extrapolated free-free emission between the detection at 1.3 and the non-detections at \SI{6.0}{\cm} in the right panel of Fig.~\ref{fig:SED_time_variability}). Despite our results, some studies (e.g., \citealt{Espaillat2019, Espaillat2022}) have not reported evidence of variability in Class II sources, so it remains unclear what makes some systems variable (and not others).

Assuming that the free-free emission in CX~Tau comes from a jet, we consider the evidence by \cite{Anglada2015}  of an almost-linear relationship between the free-free emission of a radio jet and its momentum outflow rate. Thus, the observed variability of a factor of ${\sim}20$ would imply approximately the same level of variability in $\dot{M}_\mathrm{acc}$ assuming that the jet velocity and the $\dot{M}_\mathrm{jet}/\dot{M}_\mathrm{acc}$ do not vary significantly (e.g., \citealt{Cabrit2007}). Assuming, instead, that the free-free is emitted by a photoevaporative wind, the high state at 1.3~cm would correspond to an ionizing luminosity $\Phi_\mathrm{EUV}\sim10^{40}-10^{41}$~photons~$\mathrm{s}^{-1}$ \citep{Pascucci2012}, which is considered typical in Class~II sources \citep{Coutens2019}. For variability of the order of a couple of weeks, the medium irradiated by this EUV source must be rather dense to ensure that the recombination timescale is short enough. In particular, the electron density of the ionized gas at sub-au scales would be $n>10^6\,\mathrm{cm^{-3}}$ \citep{Hollenbach2009} and this would correspond to a wind mass loss rate of ${\sim}10^{-9}\,M_\odot\,\mathrm{yr}^{-1}$ \citep{Hollenbach1994}. Another possible description for the high-intensity observation at \SI{1.3}{\cm} could be also a radio flare, a known phenomenon due to the high level of activity in  pre-main sequence stars (e.g., \citealt{Bower2003}, \citealt{Forbrich2008}, \citealt{Rivilla2015} report radio flares from the Orion nebula cluster). However, it should be noticed that radio flares often exhibit a negative or flat spectral index, indicative of non-thermal gyro-synchrotron emission, which is contrary to what we have found. These radio flares are usually associated with X-ray flaring activity. Unfortunately, no X-ray observations of CX~Tau are available for comparison. Additionally, the absence of spectra of CX Tau at other wavelengths prevents us from inspecting its accretion properties and supporting the interpretation of free-free variability. All these elements further highlight the peculiarity of the findings in CX~Tau and warrant caution when interpreting millimeter and non-simultaneous centimeter wavelength observations when evaluating the strength of a non-dust emission component.

As another potential explanation for the observed behavior, we mention the contribution from anomalous microwave emission (AME), whose presence in protoplanetary disks has been claimed by \cite{Greaves2018}. The origin of this emission is uncertain. It is thought to be the effect of electric dipole radiation from nanometric spinning dust grains \citep{DraineLazarian1998}, but the carrier is still debated. \cite{Greaves2018} discarded an origin hypothesis from polycyclic aromatic hydrocarbons, in favor of one based on nanodiamonds. The presence of AME would generate a characteristic bump at a wavelength of ${\sim}1$~cm in the spectral flux density distribution (see Fig.~1 in \citealt{Greaves_Mason_2022}) that resembles the trend of CX~Tau detections at \SI{0.9}{\mm}, \SI{1.3}{\cm}, and \SI{2.0}{\cm}. Given the uncertainties around this emission mechanism and its properties in protoplanetary disks, simultaneous multiwavelength data are needed to corroborate this scenario.

\begin{acknowledgements}

We thank the anonymous referee for the highly valuable feedback that contributed to improving this paper.
This work was partly supported by the Italian Ministero dell Istruzione, Universit\`a e Ricerca through the grant Progetti Premiali 2012 – iALMA (CUP C$52$I$13000140001$), 
by the Deutsche Forschungs-gemeinschaft (DFG, German Research Foundation) - Ref no. 325594231 FOR $2634$/$1$ TE $1024$/$1$-$1$, 
and by the DFG cluster of excellence Origins (www.origins-cluster.de). 
This project has received funding from the European Union's Horizon 2020 research and innovation programme under the Marie Sklodowska-Curie grant agreement No. 823823 (RISE DUSTBUSTERS project) and from the European Research Council (ERC) via the ERC Synergy Grant {\em ECOGAL} (grant 855130). S.F. is funded by the European Union under the European Union’s Horizon Europe Research \& Innovation Programme 101076613 (UNVEIL). G.R. acknowledges support from the Netherlands Organisation for Scientific Research (NWO, program number 016.Veni.192.233). Funded by the European Union (ERC DiscEvol, project number 101039651). Views and opinions expressed are however those of the author(s) only and do not necessarily reflect those of the European Union or the European Research Council Executive Agency. Neither the European Union nor the granting authority can be held responsible for them.

This paper makes use of the following ALMA data: ADS/JAO.ALMA\#2013.1.00426.S,and  ADS/JAO.ALMA\#2016.1.00715.S. ALMA is a partnership of ESO (representing its member states), NSF (USA) and NINS (Japan), together with NRC (Canada), MOST and ASIAA (Taiwan), and KASI (Republic of Korea), in cooperation with the Republic of Chile. The Joint ALMA Observatory is operated by ESO, AUI/NRAO and NAOJ.
\end{acknowledgements}

\bibliographystyle{aa} % style aa.bst
\bibliography{bibliography.bib} % your references Yourfile.bib

\begin{appendix}

\section{Property table and images from VLA and ALMA observations}
\label{app:images_table}

\begin{figure*}[]
    \centering
    \includegraphics[width=1.0\textwidth]{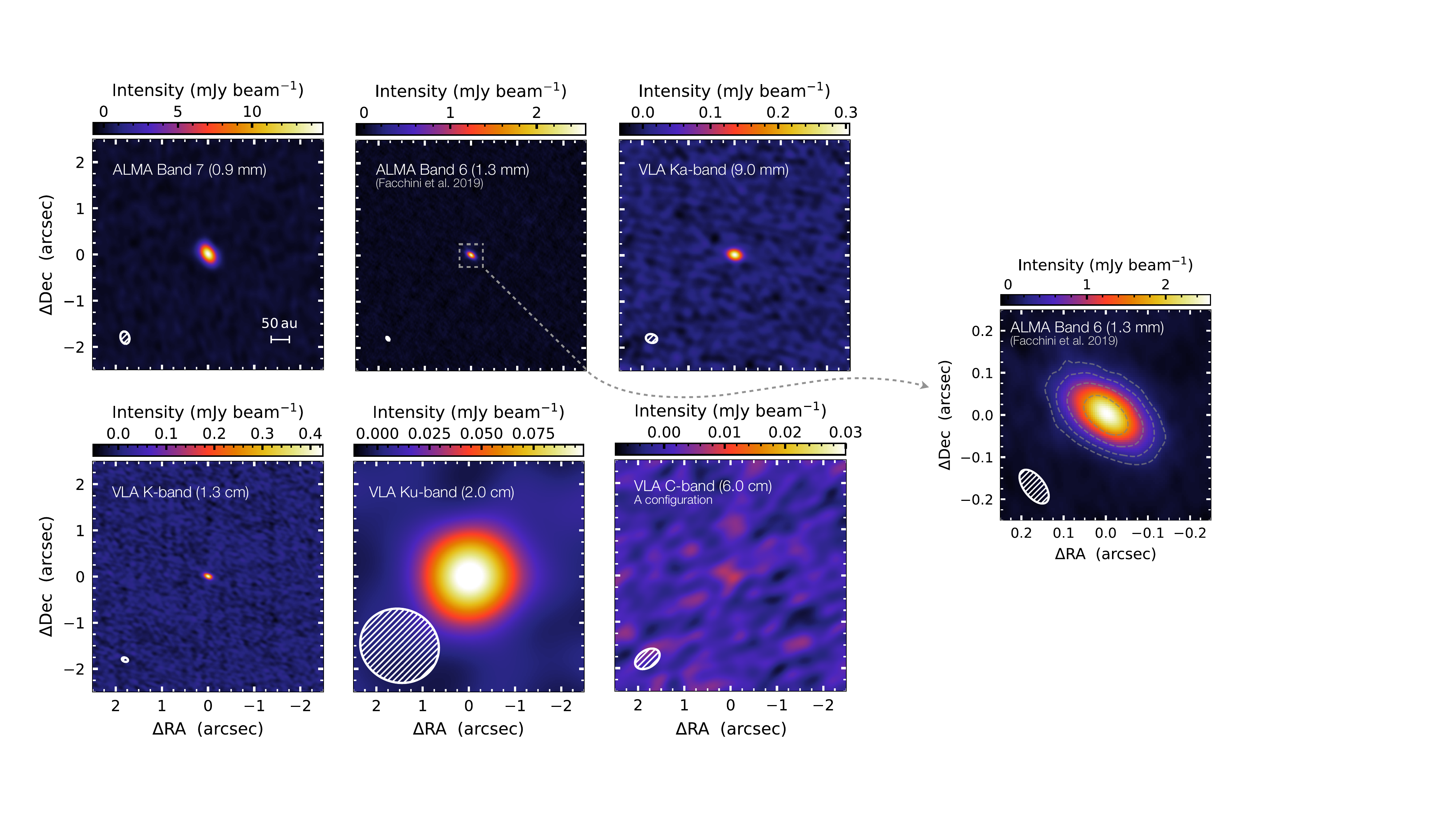}
    \caption{Continuum images of CX~Tau from ALMA and VLA observations at different wavelengths, all obtained with Briggs weighting and robust~1. The full width at half maximum (FWHM) of the synthesized beam is indicated by the ellipse in the bottom left corner of each image. The rightmost panel shows a zoomed image of the ALMA 1.3~mm observation from \cite{Facchini_19} which has the highest angular resolution. Contours indicate the [5, 10, 20, 40]$\sigma$ levels.}
    \label{fig:all_obs_images}
\end{figure*}

Table~\ref{table:all_data} provides a summary of the datasets used in this work and the derived properties.
Figure~\ref{fig:all_obs_images} shows the images from the ALMA and VLA observations of CX~Tau, along with a zoomed version of the highest angular resolution observation at \SI{1.3}{\mm} from \cite{Facchini_19}.

\begin{table*}
\caption{Parameters of CX~Tau observations at all the frequencies we considered.}             
\label{table:all_data}      
\centering      
\setlength{\tabcolsep}{5pt}
\renewcommand{\arraystretch}{3.0}
\begin{tabular}{c c c c c c c c }     
\hline\hline       
\renewcommand{\arraystretch}{1.1} \begin{tabular}{@{}c@{}} Wavelength\\ \& Observatory\\ \end{tabular}              &   \renewcommand{\arraystretch}{1.1} \begin{tabular}{@{}c@{}} Observation \\ date \\ \end{tabular} &         Beam       &  $F_\mathrm{tot}$ &       RMS         & $\Delta F_{stat}$ &  $\Delta F_{\mathrm{tot}}$ &  $R_{68\%}$    \\ 
                        &      (d/m/y)         &       (arcsec)     &         (mJy)     & (mJy beam$^{-1}$) &       (mJy)           &         (mJy)             &     (au)         \\ 
\hline                    
\renewcommand{\arraystretch}{1.1} \begin{tabular}{@{}c@{}} \tablefootmark{(a)} \SI{0.85}{\mm}\\ SMA \\ \end{tabular}         &    02/2004 - 01/2005   &        ${\sim}15$    &         25.0      &        6.0        &        6.0                     &           6.5             &    - \\

\renewcommand{\arraystretch}{1.1} \begin{tabular}{@{}c@{}c@{}} \tablefootmark{($\star$)}\SI{0.9}{\mm} \\ ALMA Band 7 \end{tabular}  &     \renewcommand{\arraystretch}{1.1} \begin{tabular}{c@{}c@{}} 
   24/07/2015 \\ 23/07/2016\\ \end{tabular} &  $0.28\times0.19$  &         21.8      &       0.12        &        0.38                   &           2.2             &  $16.9\pm0.1$ \\

\renewcommand{\arraystretch}{1.1} \begin{tabular}{@{}c@{}} \tablefootmark{(b)}   \SI{1.3}{\mm}  \\ ALMA Band 6 \end{tabular} &   \renewcommand{\arraystretch}{1.1} \begin{tabular}{@{}c@{}} C40-5: 05/11/2016 \\ C40-8: 25/09/2017 \\ \end{tabular}          &  $0.06\times0.03$  &         9.75      &       0.020       &        0.12                 &          0.98             &  $14.0\pm0.3$ \\

\renewcommand{\arraystretch}{1.1} \begin{tabular}{@{}c@{}} \tablefootmark{(c)}        \SI{1.3}{\mm}  \\ PdBI \\ \end{tabular} & 12/2010 - 02/2013           &  $0.6\times0.4$  &         9.6      &       0.2       &        0.2                   &          0.98             &  -  \\

\renewcommand{\arraystretch}{1.1} \begin{tabular}{@{}c@{}} \tablefootmark{(d)}        \SI{3.5}{\mm} \\ PdBI \end{tabular}      & 07-08/2007  &    ${\sim} 3-4$      &         1.01      &       0.13        &        0.13                  &          0.16             &  - \\
        
\renewcommand{\arraystretch}{1.1} \begin{tabular}{@{}c@{}} \tablefootmark{($\star$)}\SI{9.0}{\mm}   \\ VLA Ka-band \end{tabular}  & 12/10/2020     &  $0.25\times0.19$  &        0.304      &$7\times10^{-3}$   & $4\times10^{-3}$             &     $31\times10^{-3}$      &  - \\

\renewcommand{\arraystretch}{1.1} \begin{tabular}{@{}c@{}} \tablefootmark{($\star$)}\SI{1.3}{\cm}    \\ VLA K-band\end{tabular}& 09/04/2022      &  $0.14\times0.09$  &        0.423      &$10\times10^{-3}$  & $6\times10^{-3}$              &     $43\times10^{-3}$      &  $<9.2\pm1.5$ \\

\renewcommand{\arraystretch}{1.1} \begin{tabular}{@{}c@{}} \tablefootmark{($\star$)}\SI{2.0}{\cm}    \\ VLA Ku-band \end{tabular}         & 02-04/03/2020         &  $1.7\times1.6$    &        0.108      &$3\times10^{-3}$  & $2\times10^{-3}$               &     $11\times10^{-3}$      &  - \\

\renewcommand{\arraystretch}{1.1} \begin{tabular}{@{}c@{}} \tablefootmark{($\star$)}\SI{6.0}{\cm}   \\ VLA C-band \end{tabular}          &  C conf: 20/02/2020     &  $4.2\times3.6$    &        $<17\times10^{-3}$      &$8\times10^{-3}$  &        -         &       -      &  - \\

\renewcommand{\arraystretch}{1.1} \begin{tabular}{@{}c@{}} \tablefootmark{($\star$)}\SI{6.0}{\cm}    \\ VLA C-band \end{tabular}    &  A conf: 09-11/01/2021  &  $0.6\times0.4$    &        $<12\times10^{-3}$      &$2\times10^{-3}$  &         -         &       -     &  - \\

\hline       
\end{tabular}

\tablefoot{\tablefoottext{$\star$}{We have directly calibrated and analyzed these observations. For these data, the beam is taken from the images produced with Briggs weighting and robust~1.}
\tablefoottext{a}{}\tablefoottext{b}{}\tablefoottext{c}{}\tablefoottext{d}{}{Values from \cite{Andrews_Williams_2005}, \cite{Facchini_19}, \cite{Pietu2014}, and \cite{Ricci10}, respectively.}
Systematic calibration errors are assumed to be $10\%$ of the total flux density, apart from VLA C-Band observations where we used a value of $5\%$. ALMA calibration errors are based on \url{https://almascience.eso.org/documents-and-tools/cycle9/alma-technical-handbook}, while VLA calibration errors follow the indications by NRAO at \url{https://science.nrao.edu/facilities/vla/docs/manuals/oss/performance/fdscale}. For the remaining observations, we refer to the respective papers.}
\end{table*}

\section{Checks and treatment of the 1.3~cm VLA data}
\label{Kband_checks}
\begin{figure*}[]
\centering
\includegraphics[width=\hsize]{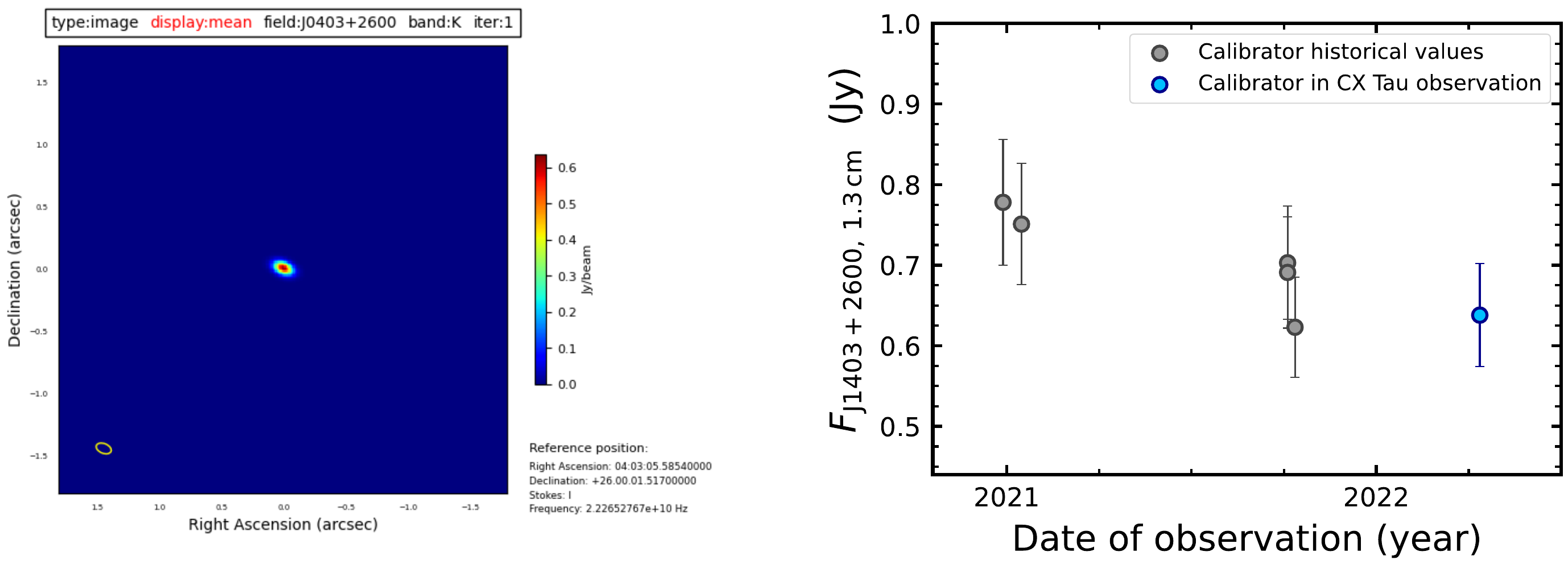}
\caption{Checks on the gain calibrator J0403+2600 during the CX~Tau observation at 1.3~cm. \textit{Left panel}: Gain calibrator J0403+2600 image at 1.3~cm (K band) from the weblog of the CX~Tau observation at the same wavelength.  \textit{Right panel}: Time variability of the 1.3~cm flux density from the gain calibrator J0403+2600.}
\label{fig:J0403+2600_checks}
\end{figure*}

To address the unexpectedly high value of the 1.3~cm VLA data, we present evidence in this section to validate the data collection process and the self-calibration. We carefully examine the gain calibrator J0403+2600 to eliminate the possibility of an incorrect flux transfer that may have influenced the measurement of CX Tau's flux. The left panel of Fig. \ref{fig:J0403+2600_checks} presents the image of J0403+2600 from the observation weblog, where nothing seems to be worth of notice. The right panel of Fig. \ref{fig:J0403+2600_checks} shows the time variability of the flux of the calibrator, comparing the flux value measure in the CX Tau's execution block to the flux historical values from the source. During the CX Tau observation, the flux value of J0403+2600 is perfectly in line with the historical values. Therefore, the anomalously high value of the 1.3~cm data cannot be explained with an incorrect flux calibration.

We conducted a meticulous four-step phase-only self-calibration procedure. Prior to self-calibration, the RMS was measured at 9.5~$mu$Jy/beam, and the flux density from the CX~Tau disk was 70~$mu$Jy, resulting in a peak S/N of 5.1. Throughout the self-calibration process, the RMS remained unchanged, while CX~Tau's flux density steadily increased. The following are the details of each self-calibration step, along with the parameters used for the \texttt{gaincal} task in \texttt{CASA}:
\begin{enumerate}
    \item We solved for shifts in polarization only with \texttt{gaintype='G'}, \texttt{combine='scan, spw'}, and \texttt{minsnr=3}. As a result, the source flux density increased to 100~$mu$Jy.
    \item We corrected shifts only between spectral windows with \texttt{gaintype='T'}, \texttt{combine='scan'}, and \texttt{minsnr=3}. CX~Tau's flux density reached 160~$mu$Jy.
    \item We obtained solutions for both polarization and spectral window shifts with \texttt{gaintype='G'}, \texttt{combine='scan'}, and \texttt{minsnr=3}. This caused the flux density to rise to 260~$mu$Jy.
    \item We repeated the same command in the fourth step but with a lower threshold for the S/N of the solutions (\texttt{gaintype='G'}, \texttt{combine='scan'}, \texttt{minsnr=2}).
\end{enumerate}
At the conclusion of the process, the RMS remained consistent at 9.5~$mu$Jy/beam, while CX~Tau's flux density increased to 420~$mu$Jy, resulting in a peak S/N of 40. It is important to notice that such RMS value is in excellent agreement with the theoretical noise predicted by the VLA exposure calculator\footnote{\url{https://obs.vla.nrao.edu/ect/}} which is 9.0~$mu$Jy/beam. From the self-calibration procedure, we can infer that the increase in flux density, coupled with the unchanged RMS, is a result of phase decoherence that dispersed the flux from the central source throughout the field of view.  As CX Tau's flux was not notably high, this effect had minimal impact on the RMS. Consequently, self-calibration successfully gathered and refocused the dispersed flux, leading to an increased integrated flux of the central source.

\section{\texttt{galario} fits}
\label{app:galario_resuts}

\begin{figure*}[]
    \centering
    \includegraphics[width=1\textwidth]{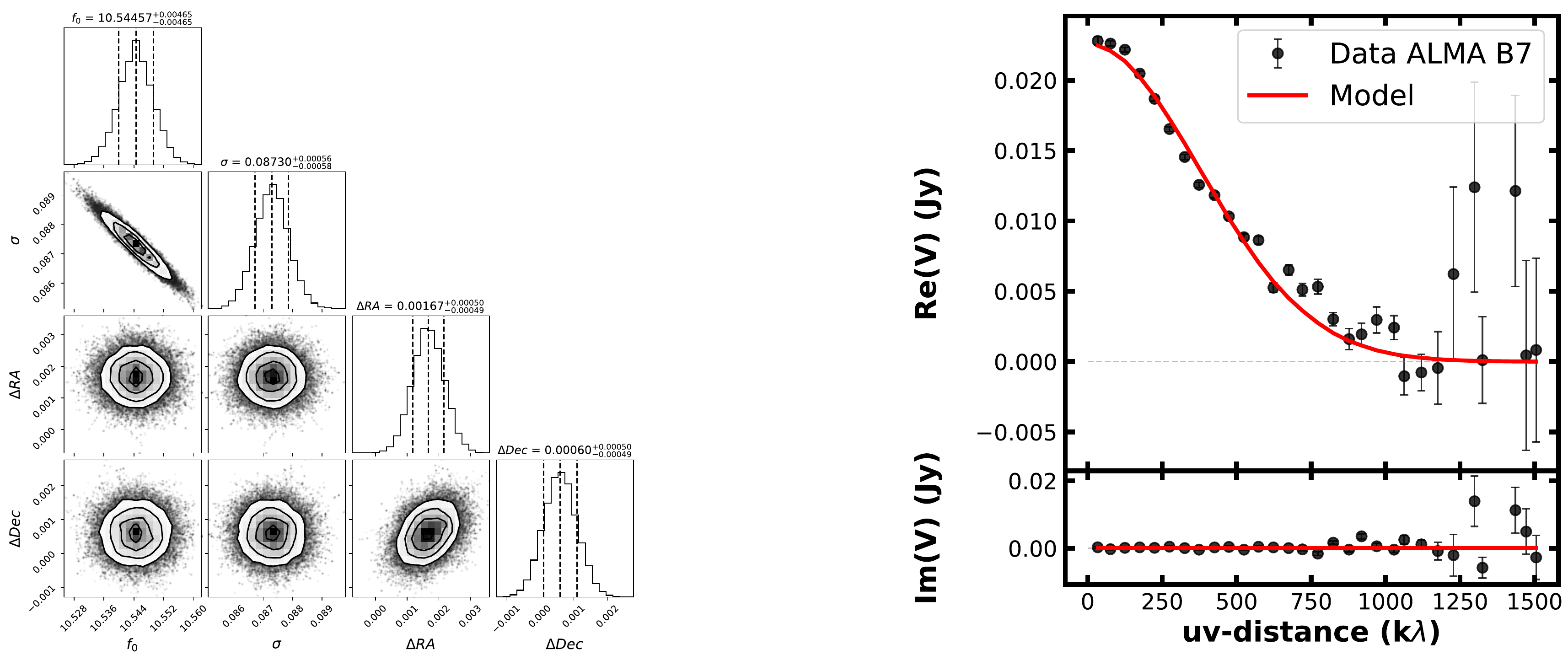}
    \caption{Results of the \texttt{galario} fit with a Gaussian model on ALMA Band 7 (\SI{0.9}{\mm}) data.  \textit{Left panel}: Corner plot of the MCMC run.\textit{ Right panel}: Recentered and deprojected visibilities binned in 50k$\lambda$ intervals and the best-fit model. Error bars are at $1\sigma$.}
    \label{fig:ALMAB7_galario_results}
\end{figure*}

\begin{figure*}[]
    \centering
    \includegraphics[width=1\textwidth]{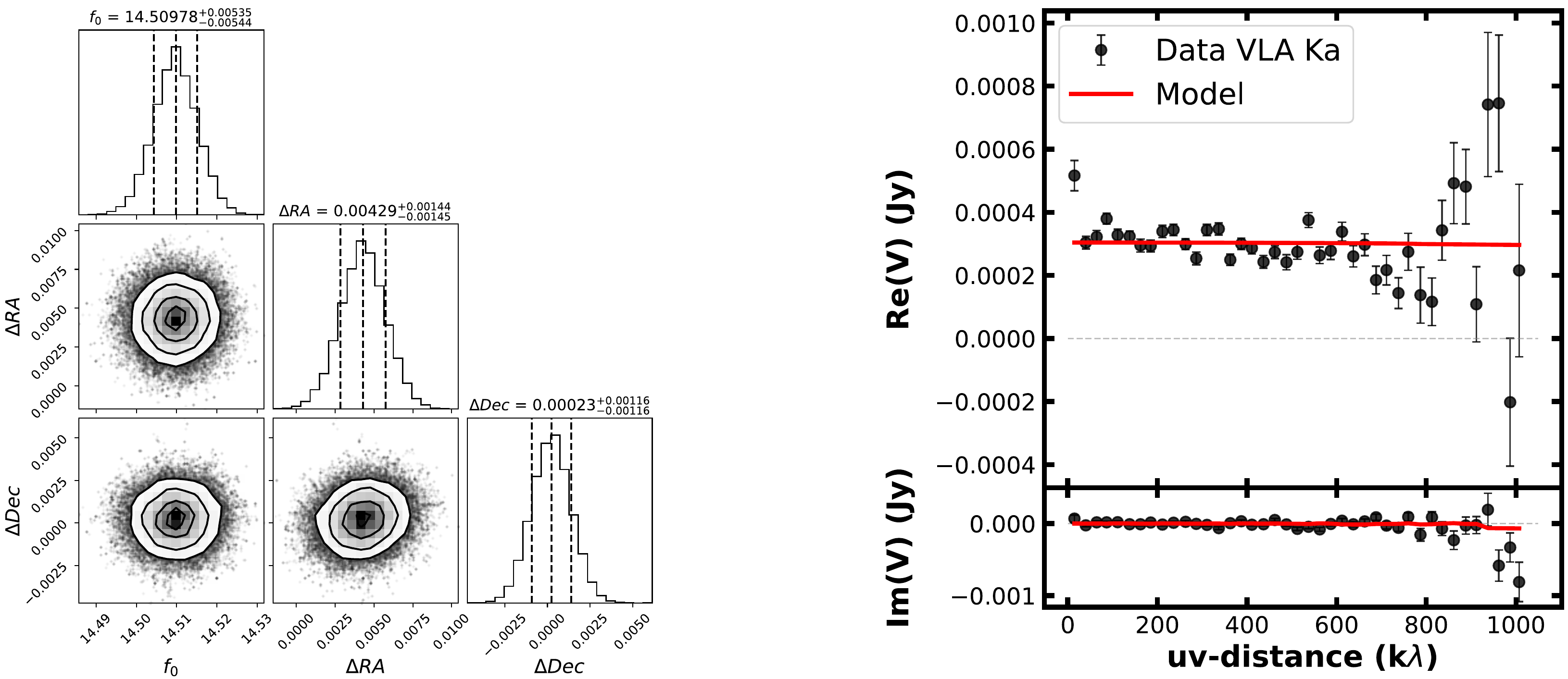}
    \caption{Results of the \texttt{galario} fit with a point source model on the VLA Ka-band (\SI{9.0}{\mm}) data.  \textit{Left panel:} Corner plot of the MCMC run.\textit{ Right panel:}  Recentered and deprojected visibilities binned in 25k$\lambda$ intervals and the best-fit model. Error bars are at $1\sigma$.}
    \label{fig:VLA_Ka_galario_results}
\end{figure*}

\begin{figure*}[]
    \centering
    \includegraphics[width=1\textwidth]{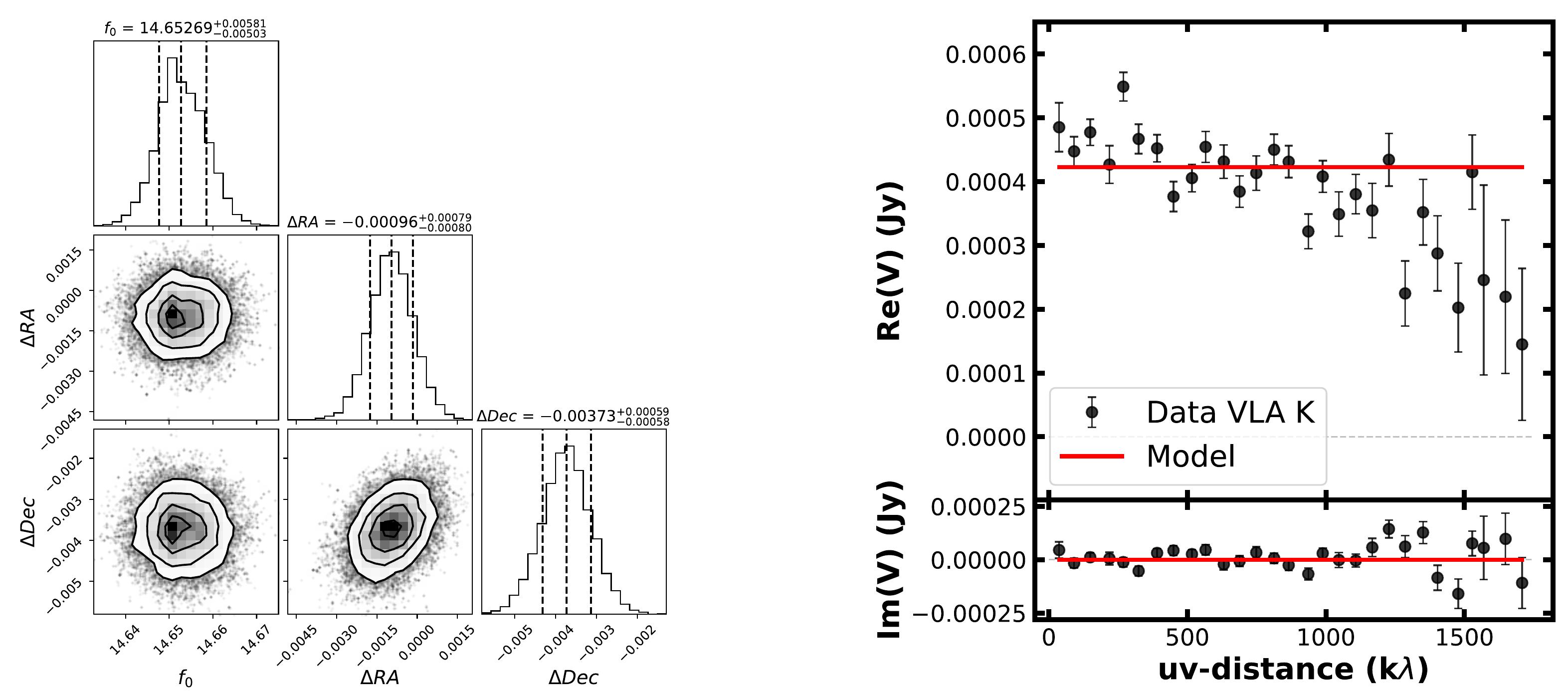}
    \caption{Results of the \texttt{galario} fit with a point source model on the VLA K-band (\SI{1.3}{\cm}) data. \textit{Left panel:} Corner plot of
the MCMC run.\textit{ Right panel:}  Recentered and deprojected visibilities binned in 60k$\lambda$ intervals and the best-fit model. Error bars are at $1\sigma$.}
    \label{fig:VLA_K_galario_results}
\end{figure*}

\begin{figure*}[]
    \centering
    \includegraphics[width=1\textwidth]{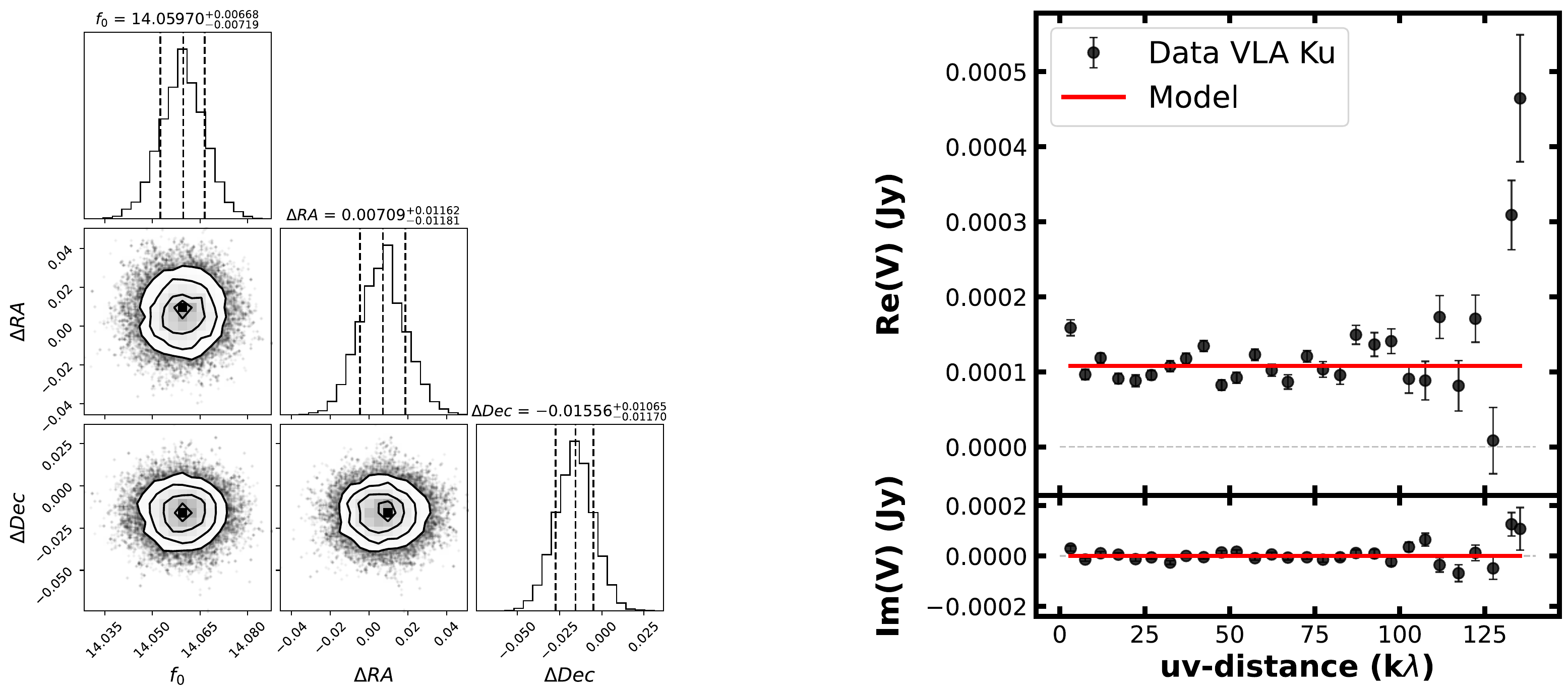}
    \caption{Results of the \texttt{galario} fit with a point source model on the VLA Ku-band (\SI{2.0}{\cm}) data.  \textit{Left panel:} Corner plot of
the MCMC run.\textit{ Right panel:}  Recentered and deprojected visibilities binned in 5k$\lambda$ intervals and the best-fit model. Error bars are at $1\sigma$.}
    \label{fig:VLA_Ku_galario_results}
\end{figure*}

To retrieve the flux density, and the radial extent in the case of a resolved observation, we fit the data using the code \texttt{galario} \cite{Tazzari2018_galario}. It works by assuming a 1D or 2D model of the emission in the image plane and performing a Fourier transform to obtain the synthetic visibilities at the same $uv$-points of the observations. The best-fit model is found as the one that minimizes the $\chi^2$ by sampling the parameter space with an MCMC approach using the \texttt{emcee} package \citep{emcee}. We employed uniform priors and the intensity normalization factor $f_0$ was sampled logarithmically: $\log_{10}(f_0/(\mathrm{Jy/sr})) \in [0, 30]$, $\Delta\mathrm{RA}\in[-2, \SI{2}{\arcsec}]$, $\Delta\mathrm{Dec}\in[-2, \SI{2}{\arcsec}]$, and, for the fit of ALMA \SI{0.9}{\mm} observation, also $\sigma\in[0, \SI{0.2}{\arcsec}]$. In each run, we used 100 walkers that were well converged after ${\sim}1000$ steps. Corner plots along with deprojected visibilities and best-fit model for data from ALMA Band~7 (\SI{0.9}{\mm}) and VLA Ka (\SI{9.0}{\mm}), K (\SI{1.3}{\cm}), and Ku (\SI{2.0}{\mm}) bands are shown in Figs.~\ref{fig:ALMAB7_galario_results}, \ref{fig:VLA_Ka_galario_results}, \ref{fig:VLA_K_galario_results}, and \ref{fig:VLA_Ku_galario_results}.

\section{Intraband spectral indices for VLA observations}
\label{app:intraband_spectral_indices}

\begin{figure}[]
\centering
\includegraphics[width=\hsize]{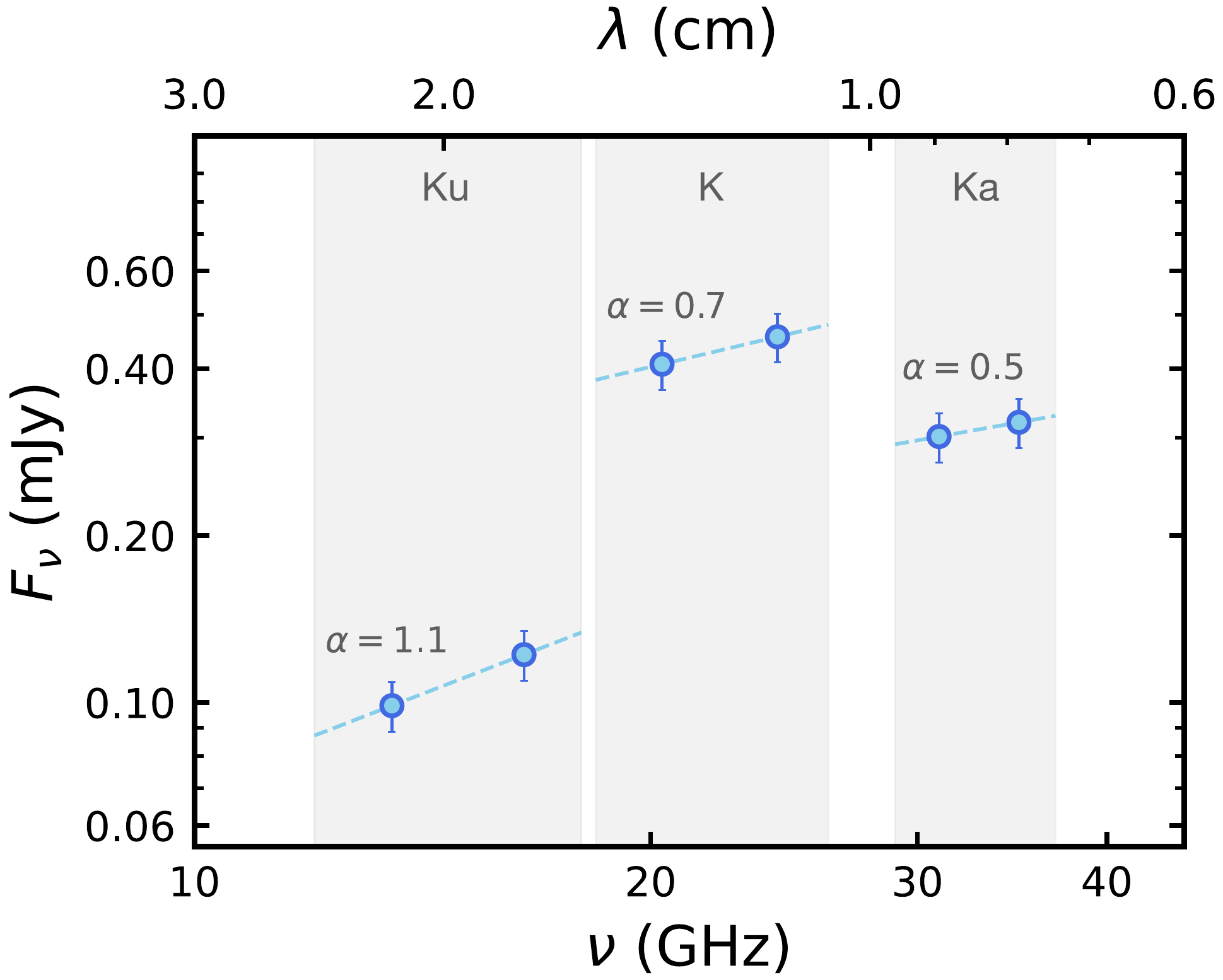}
\caption{Spectral flux density distribution of CX~Tau zoomed in on the VLA data in Ka (\SI{9.0}{\mm}), K (\SI{1.3}{\cm}), and Ku (\SI{2.0}{\mm}) bands. Each observation is split into high and low frequencies and then fitted to obtain the intraband spectral indices reported in the plot.}
\label{fig:intrband_spectral_index}
\end{figure}

To prove that the VLA detections in Ka (9.0 mm), K (1.3 cm), and Ku (2.0 mm) bands are all consistent with free-free emission, despite the high-amplitude variability, we show in Fig.~\ref{fig:intrband_spectral_index} their intraband fluxes. We split each observation into two halves between high and low frequencies and extracted the flux densities. Then, we used a power law to fit the two intensities retrieved from each band and obtained the intraband spectral indices values $\alpha=0.5, \, 0.7,  \, 1.1$ for the data from the Ka, K, and Ku bands, respectively.

\section{Upper limit for radius estimate from K-band data}
\label{app:upper_lim_Kband}

\begin{figure}[]
\centering
\includegraphics[width=\hsize]{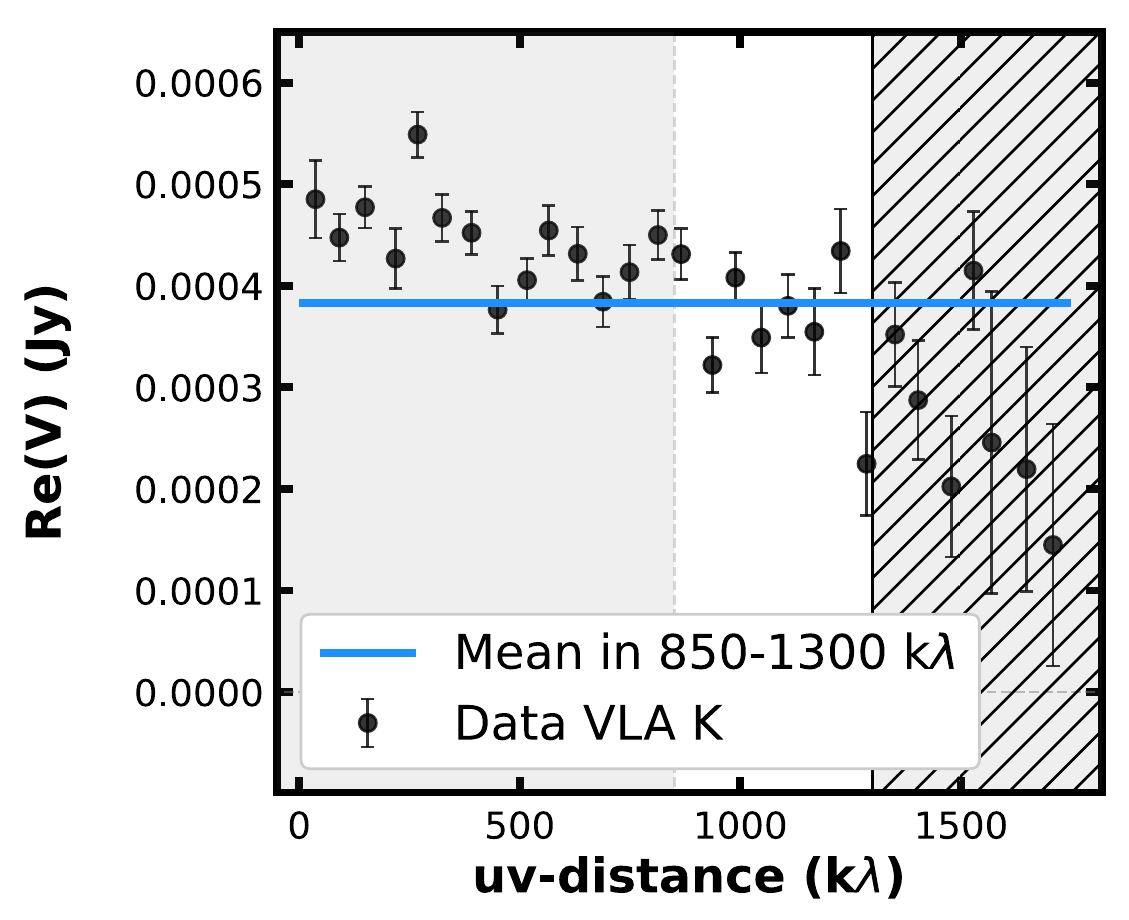}
\caption{Real part of the visibilities from VLA K-band (\SI{1.3}{\cm}) data binned in 60k$\lambda$ intervals and overlaid blue line indicating their mean value between 850 and 1300k$\lambda$ (range with white background). The hatched area contains the visibilities with a $uv$-distance greater than 1300k$\lambda$ which have been excluded in the radius evaluation due to high noise.}
\label{fig:CXTau_K_cut_rescale}
\end{figure}

\begin{figure*}[]
    \centering
    \includegraphics[width=1\textwidth]{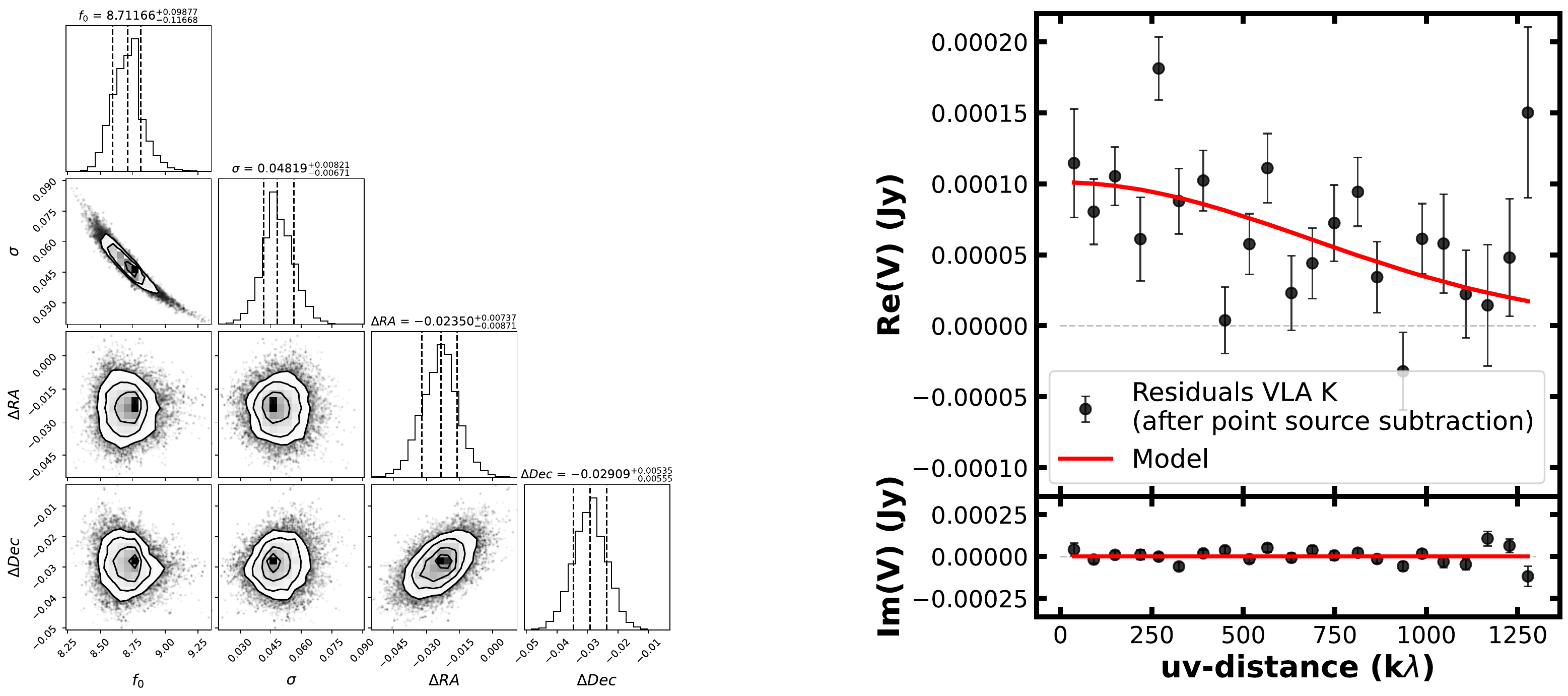}
    \caption{Results of the \texttt{galario} fit with a Gaussian model on VLA K-band (\SI{1.3}{\cm}) residual data, after subtracting  from the real part of the visibilities their mean values in the range of $uv$-distance [850, 1300]k$\lambda$ and neglecting the visibilities over 1300k$\lambda$.  \textit{Left panel}: Corner plot of the MCMC run. \textit{Right panel:}  Recentered and deprojected visibilities binned in 60k$\lambda$ intervals and best-fit model. Error bars are at $1\sigma$.}
    \label{fig:VLA_K_galario_results_cut_rescaled}
\end{figure*}

The right panel of Fig.~\ref{fig:VLA_K_galario_results} shows that the K-band (\SI{1.3}{\cm}) observation is mostly non-resolved. However, a slight downward trend in the visibility profile is visible, possibly hinting at a resolved component in the data. To extract this component, we first excluded the visibilities at a $uv$-distance greater than 1300k$\lambda$ as they are dominated by noise. Then, we considered only the visibilities in a range of $uv$-distances where they appear the flattest and we chose the interval between 850 and 1300k$\lambda$. We  computed the average of the  real parts of the visibilities and obtained a flux density of \SI{0.37}{\mathrm{mJy}}, equivalent to the free-free only estimate obtained by subtracting the extrapolated dust emission from the total flux density value at \SI{1.3}{\cm} (indicated by the yellow circle at this wavelength in the left panel of Fig.~\ref{fig:SED_time_variability}).  This procedure is visualized in Fig.~\ref{fig:CXTau_K_cut_rescale}. We subtracted the computed  intensity value to the real parts of the visibilities having a $uv$-distance within 1300k$\lambda$ getting as a result the visibility profile of the resolved dust emission component in the observation. To retrieve a dust size, we fit these visibilities with \texttt{galario} using a Gaussian model profile and the same parameters used in the fit of ALMA \SI{0.9}{\mm} data (see Appendix~\ref{app:galario_resuts}). The converged fit  and the results are shown in Fig~\ref{fig:VLA_K_galario_results_cut_rescaled}.

\end{appendix}

\end{document}